\documentclass[12pt,twoside,a4paper]{article}

\pdfoutput=1

\usepackage[usenames]{color}
\usepackage{lscape}
\usepackage{amssymb}
\usepackage{rotating}
\usepackage[T1]{fontenc}
\usepackage{graphicx}
\usepackage{fancyhdr}
\usepackage{amsmath}
\usepackage{ae}
\usepackage{aecompl}
\usepackage{hyperref}

\def\nn{\nonumber \\}

\newcommand{\be}{\begin{equation}}
\newcommand{\ee}{\end{equation}}
\newcommand{\ba}{\begin{eqnarray}}
\newcommand{\ea}{\end{eqnarray}}

\newlength{\dinwidth}
\newlength{\dinmargin}
\begin{document}

\thispagestyle{empty}

\vspace*{1cm}

\centerline{\Large\bf Interpreting the 750 GeV diphoton excess in minimal    } 
 \centerline{\Large\bf extensions of Two-Higgs-Doublet models }

\vspace*{15mm}

\centerline{Marcin Badziak }
\vspace*{5mm}

\centerline{\em Institute of Theoretical Physics,
Faculty of Physics, University of Warsaw} 
\centerline{\em ul.~Pasteura 5, PL--02--093 Warsaw, Poland}

\vskip 1cm

\centerline{\bf Abstract}

It is shown that the 750 GeV diphoton excess can be explained in extensions of Two-Higgs-Doublet Models that do not involve large multiplicities of
new electromagnetically charged states. The key observation is that at moderate and large $\tan\beta$ the total decay width of the 750 GeV Higgs is
strongly reduced as compared to the Standard Model. This allows for much more economical choices of new states that enhance the diphoton signal to fit
the data.
In particular, it is shown that one family of vector-like quarks and leptons with SM charges is enough to explain the 750 GeV diphoton excess.
Moreover, such charge assignment can keep the 125 GeV Higgs signal rates exactly at the SM values. The
scenario can interpret the diphoton excess provided that the total decay width of a hypothetical resonance that would be measured at the LHC
turns out to not exceed few GeV. 

\vskip 3mm

\newpage

\section{Introduction}

The ATLAS and CMS collaborations reported recently an excess in the diphoton mass distribution around 750 GeV \cite{ATLAS750,CMS750}. Local
significances of these excesses are somewhat above 3$\sigma$ at ATLAS and slightly less than $3\sigma$ at CMS. While global significance of this
excess is not yet large enough to celebrate discovery of New Physics, it is the most significant excess observed simultaneously at ATLAS and CMS in searches
for New Physics at the LHC so far. Thus, it is tempting to interpret this signal in extensions of the  Standard Model (SM).

There are many ways how to explain the 750 GeV excess by New Physics \cite{Giudice750}. Among candidates for a new resonance there are singlets
coupled to vector-like fermions \cite{Buttazzo:2015txu}-\cite{Chao:2015ttq}, composite states \cite{Harigaya:2015ezk}-\cite{Hernandez:2015ywg}, 
states originating from reduction of extra dimensions \cite{Cox:2015ckc}-\cite{Ahmed:2015uqt}, axions \cite{Pilaftsis:2015ycr}-\cite{Higaki:2015jag}
or sgoldstinos \cite{Petersson:2015mkr}-\cite{Bellazzini:2015nxw}.\footnote{It has also been suggested that the excess may not originate from a 750
GeV resonance \cite{Cho:2015nxy}. } Some authors speculate also on a possible link of this new resonance to a dark
matter particle \cite{Mambrini:2015wyu}-\cite{Bauer:2015boy}. Here, we assume that the 750 GeV diphoton excess is due to new Higgs boson(s) in
Two-Higgs-Doublet Model (2HDM) \cite{2HDMreview}. Such interpretations of the diphoton signal were already presented in
Refs.~\cite{750_Djouadi,Becirevic:2015fmu,Han:2015qqj}. In those articles the main focus was on small values of $\tan\beta$  with dominant
contribution to production of a 750 GeV states in gluon fusion coming from a top quark loop. It has been shown, however, that in order to fit the
diphoton signal 2HDM must be extended by additional new states with large multiplicities and/or large exotic electromagnetic charges.

In the present paper we investigate a possibility to fit the 750 GeV diphoton excess in extensions of 2HDMs with moderate and large $\tan\beta$. At
first sight, it might seem to be not a  good choice of parameter space because at large $\tan\beta$ top quark contribution to gluon
fusion is strongly suppressed. However, since new states have to be added anyway to 2HDMs to enhance 750 GeV Higgs decays to diphotons it is
reasonable  to assume that these new states also carry colour charge and contribute to the 750 GeV Higgs production via gluon fusion. In such a
case top quark contribution to gluon fusion is no longer necessary and $\tan\beta$ can be large. The main advantage of large $\tan\beta$ is that the
total decay width of the 750 GeV Higgs is suppressed in this regime. This
allows for much smaller diphoton decay width of the 750 GeV Higgs to explain the excess. Moreover, if the excess is due to
narrow resonance produced in gluon
fusion, preferred signal rate of this resonance is about 6 fb, as compared to 11 fb for a resonance with total decay width of 45 GeV
\cite{Falkowski750}. Due to larger diphoton signal rate  the wide resonance hypothesis is in bigger tension with LHC run-1 data
\cite{Falkowski750} (see also Ref.~\cite{Rolbiecki750pheno}). On the other hand, in the narrow resonance hypothesis, the best-fit point from 13 TeV
data is consistent with constraints from the run-1 data. Nevertheless, the best-fit point in a global fit to all diphoton data shifts downwards to
about 3 fb. 

We investigate possible
size of the suppression of the total decay width of the 750 GeV Higgses in Type-I and Type-II 2HDM and show that it is large enough to fit the
diphoton excess with rather small multiplicities of new particles. In particular, we demonstrate that one family of vector-like quarks and leptons with
SM charges is enough to explain the 750 GeV diphoton excess. By construction, this scenario can interpret the diphoton excess provided that the total
decay width of a hypothetical resonance that would be measured at the LHC
turns out to not exceed few GeV.  

\section{Suppression of a Higgs total decay width in 2HDMs and enhanced 750 GeV diphoton signal}

The total decay width of a 750 GeV Higgs in the SM is about 247 GeV \cite{Xwgr}. The main decay channels are into $WW$, $ZZ$ and $t\bar{t}$ with the
corresponding branching ratios of about 59\%, 29\% and 12\%, respectively. As a consequence of large total decay width, BR$(H\to\gamma\gamma)$ is only
$2\times10^{-7}$. Since the SM production cross-section for the 750 GeV Higgs, dominated by gluon fusion rate, is about 0.74 pb \cite{Xwgr}, it is
clear that if the 750 GeV resonance is a Higgs it must have totally different properties than in the SM. 

In 2HDMs there are three physical neutral Higgs bosons, two CP-even and one CP-odd, that originate from two Higgs doublets, $H_u$ and $H_d$. Two
important parameters of this class of models are $\tan\beta=v_u/v_d$, the ratio of vacuum expectation values of the doublet neutral components,
$H_u^0$ and $H_d^0$, and angle $\alpha$ which parametrizes the mixing between the two CP-even states:
\begin{equation}
H_u^0=\cos\alpha h + \sin\alpha H \,, \qquad  H_d^0=-\sin\alpha h + \cos\alpha H \,.
\end{equation}
In the present work, we identify $h$ with the 125 GeV Higgs, while $H$ is a candidate for the 750 GeV resonance. We  focus on the so-called
alignment limit $\alpha=\beta-\pi/2$ \cite{2HDMalignment}. In such a case $h$ has exactly the same couplings as the SM Higgs while $H$ couples to the
SM fermions but not to the gauge bosons. This is motivated, in part,  by the fact that the LHC 125 GeV Higgs data agree quite well with the SM
prediction \cite{Higgsdata}. More importantly, in the alignment limit the total decay width of $H$ is generically much smaller than in the SM. In
particular, for $\tan\beta=1$, when the $H$ couplings to the SM fermions are the same as in the SM, the total decay width is about 30 GeV. Similar
decay width has CP-odd scalar, which has the same couplings to SM particles as $H$ in the alignment limit. In spite of vanishing couplings to gauge
bosons, the branching ratios of $H$ and $A$ to diphoton are of order $10^{-5}$, much too small to explain the 750 GeV excess.

In the most widely studied Type-I and Type-II 2HDMs, the correct magnitude of the 750 GeV diphoton signal could be, in principle, adjusted by choosing
appropriately small value of $\tan\beta$. This is because the effective gluon coupling of $H/A$ is proportional to the coupling to top quark which is
rescaled by a
factor $1/\tan\beta$, as compared to the SM. However, such possibility is experimentally excluded since $t\bar{t}$ production from $H/A$ decays would
be too large.

\begin{figure}[t]
\center
\includegraphics[width=0.48\textwidth]{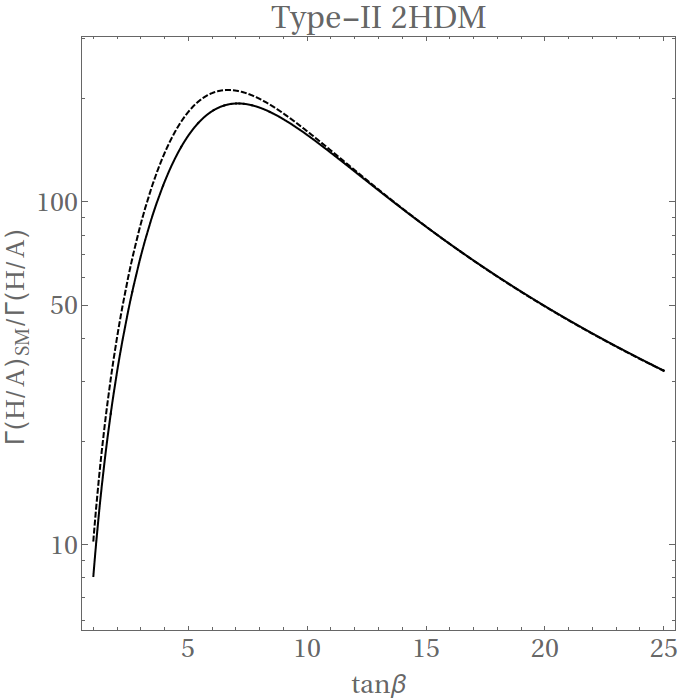}
\includegraphics[width=0.48\textwidth]{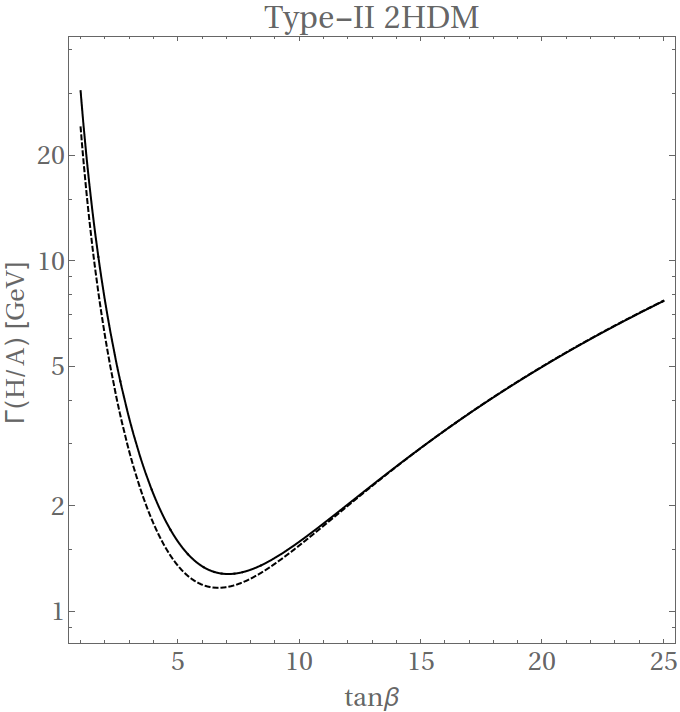}
\caption{Left panel: Enhancement of the total decay width of the 750 GeV CP-even (solid lines) and CP-odd (dashed lines) Higgs in Type-II 
 2HDM in the alignment limit $\alpha=\beta-\pi/2$, with respect to the 750 GeV SM Higgs. Right panel: the total decay width in GeV in the same case
as in the left panel. Difference between the CP-even and CP-odd Higgs comes from a different phase space suppression in $H/A\to t\bar{t}$.
}
\label{fig:decaywidth_II}
\end{figure}

The remaining possibility is to assume that there exist new electromagnetically charged particles that modify $\Gamma(H/A\to\gamma\gamma)$. In
Ref.~\cite{750_Djouadi} it was shown that it is indeed possible to fit the 750 GeV excess using decays of degenerate $H$ and $A$  to $\gamma\gamma$
enhanced by vector-like leptons. However, in such a case the price to pay is very high multiplicity of vector-like leptons. Moreover, in order not to
spoil the 125 GeV Higgs decays into photons fine cancellation in the amplitude between the contributions from different vector-like leptons is
required.
In an explicit example presented in Ref.~\cite{750_Djouadi} $\tan\beta=1$ was used, for which the model is at the verge of exclusion by the LHC
searches for $H\to t\bar{t}$.

We focus instead on larger values of $\tan\beta$ since they allow to reduce $\Gamma(H/A\to t\bar{t})$, hence also the total decay width. The
reduction of the $H/A$ couplings to top quarks results also in decrease of the gluon fusion production cross-section. Therefore, in this case new
particles should exist that carry colour charge that are responsible for large enough production cross-section of $H/A$, however, as we will see with
much smaller multiplicity than for $\tan\beta=1$. Since couplings of $H$ and $A$ to bottom quarks are different in type-I and type-II 2HDMs we
discuss these models separately in the following subsections.

\subsection{Type-II 2HDM}

In type-II 2HDM, in which the Higgs sector is that of MSSM, the couplings of $H$ and $A$ to bottom quarks are proportional to $\tan\beta$. 
For the SM Higgs with mass of 750 GeV the decay width into top quarks is about 2900 times larger than that into bottom quarks \cite{Xwgr}. This
implies in Type-II 2HDM that those decay widths equalize at $\tan\beta\approx7.3$. At this value of $\tan\beta$ the total decay width of $H$ is
minimized and equals around 1 GeV, as can be seen in the left panel of Figure~\ref{fig:decaywidth_II}. Hence, it is smaller by more than two
orders of
magnitude than for the SM Higgs with the same mass, and by a factor of 30 as compared to the $\tan\beta=1$ case. However, BR$(H\to\gamma\gamma)$ is
not enhanced because reduced $H$ coupling to top quarks
reduces also the top contribution to $\Gamma(H\to\gamma\gamma)$. The same applies to the decays of the CP-odd Higgs. Moreover, the cross-section
for production of $H$ and $A$ via gluon fusion is suppressed by $1/(\tan\beta)^2$. Nevertheless, this can be fixed by introducing new particles that
are both electromagnetically and coloured charged. The 2HDMs do not have coloured particles in the spectrum but they can be treated as simplified
models of some more complete models where such particles are present. For the sake of demonstration, we consider the model proposed in
Ref.~\cite{750_Djouadi} but with both vector-like quarks and leptons.\footnote{Phenomenology of vector-like fermions and their impact on Higgs production and
decays were
investigated e.g. in Refs.~\cite{Joglekar:2012vc}-\cite{Ellis:2014dza}.} The key feature of that model is that up-type
and down-type vector-like fermions couple to $H_u$ and $H_d$, respectively. In consequence,  contributions from different types of vector-like
fermions to the amplitude for Higgs decaying to photons/gluons have different dependence on the mixing angle $\alpha$:\cite{750_Djouadi}
\begin{align}
\label{Agg}
 \mathcal{A}_{\rm VLF}^{\Phi}(gg)&\sim \mathcal{A}_{\rm top/bottom}^{\Phi}(gg)
+ \sum_i^n\left[\sin\alpha\, \frac{vg_{u_i}}{m_{u_i}}A_{1/2}^{\Phi}(\tau_{u_i}) + \cos\alpha\, 
\frac{vg_{d_i}}{m_{d_i}}A_{1/2}^{\Phi}(\tau_{d_i})
\right] \,, \\
\label{Agamgam}
 \mathcal{A}_{\rm VLF}^{\Phi}(\gamma\gamma)&\sim \mathcal{A}_{\rm top/bottom/W}^{\Phi}(\gamma\gamma)
+ \sum_i^n\left[\sin\alpha\,  N_c^{u_i}\frac{vg_{u_i}Q_{u_i}^2}{m_{u_i}}A_{1/2}^{\Phi}(\tau_{u_i}) +
\cos\alpha N_c^{d_i}\, \frac{vg_{d_i}Q_{d_i}^2}{m_{d_i}}A_{1/2}^{\Phi}(\tau_{d_i}) + \right. \nn
&+ \left. \sin\alpha\,  \frac{vg_{\nu_i}Q_{\nu_i}^2}{m_{\nu_i}}A_{1/2}^{\Phi}(\tau_{\nu_i}) +
\cos\alpha\, \frac{vg_{l_i}Q_{l_i}^2}{m_{l_i}}A_{1/2}^{\Phi}(\tau_{l_i})
\right]
\end{align}
for $\Phi=H,A$ in the alignment limit ($\alpha=\beta-\pi/2$), while for $\Phi=h$  $\sin\alpha \to \cos\alpha$ and  $\cos\alpha \to -\sin\alpha$
should be substituted in the
above formulae. In the above formula $\nu_i$ ($l_i$) correspond to up-type (down-type) vector-like leptons. The form factors for spin-$1/2$ fermions
$A_{1/2}^{\Phi}(\tau_i)$ with $\tau=M_{\Phi}^2/(4m_i^2)$, as well as SM contributions from
top, bottom and $W$ boson can be found e.g. in Ref.~\cite{Djouadireview}. The form factors are maximized for $\tau\approx1$, in the limit $\tau\to0$
they approach values of order one, while in the limit $\tau\to\infty$ they go to zero (but rather slowly). Moreover, the form factors are typically
slightly larger for CP-odd than for CP-even Higgses. It is important to note that for all Higgses top quark dominates the contribution to gluon fusion
from the SM particles. While in the $h\to\gamma\gamma$ amplitude, dominant W boson contribution interferes destructively with subdominant (but
non-negligible) top contribution.

From the perspective of the diphoton excess the most interesting region is the one with $\tan\beta$ around 6 to 8, where the total decay width of $H$
and $A$ is minimal.
In this region the contributions from SM particles to gluon fusion and $\gamma\gamma$ amplitude are strongly suppressed. Therefore, in order to
explain the 750 GeV diphoton signal some of the new particles must carry colour and electromagnetic charge. However, due to suppressed total decay
width only few new particles are required, in contrast to the $\tan\beta=1$ case considered in Ref.~\cite{750_Djouadi}. In what follows we assume that
there is only one family of vector-like quarks and leptons with the same pattern of charges as the SM fermions:
\begin{equation}
\label{vectorlikeset}
 \binom{t'}{b'}_{L/R}, t^{''}_{L/R}, b^{''}_{L/R} , \binom{\nu'}{l'}_{L/R}, \nu^{''}_{L/R}, l^{''}_{L/R} \,.
\end{equation}
We assume that the mixing between the vector-like fermions and the SM fermions is negligible. As emphasized in Ref.~\cite{750_Djouadi}, it is crucial
to introduce both $'$ and $''$ states to have gauge invariant Yukawa interactions for the vector-like fermions. Hence, $n=2$ should be used in the
formulae \eqref{Agg}-\eqref{Agamgam} for the amplitudes. In these formulae $g_i$ are the Yukawa couplings of the vector-like fermions in the mass
basis. They are functions of the Yukawa couplings and explicit mass terms for the vector-like fermions in the interactions basis. For simplicity, we
assume that $g_i$ are free parameters. The key feature of this model is a different $\alpha$-dependence of the vector-like contributions to
the gluon fusion and $\gamma\gamma$ amplitudes for $h$ and $H/A$. This implies that if contributions from vector-like up-type and down-type fermions 
interfere constructively in the $H/A$ amplitudes, in the $h$ amplitudes they interfere destructively. In general, it is not possible to exactly
cancel vector-like fermion contributions simultaneously in $h\to\gamma\gamma$ and $h\to gg$ amplitudes. However, it follows from
eqs.~\eqref{Agg}-\eqref{Agamgam} with $\sin\alpha \to \cos\alpha$ and  $\cos\alpha \to -\sin\alpha$ that such cancellation is possible for some
combinations of masses, couplings and charges if both vector-like quarks and leptons couple to the Higgs. In order to better illustrate this fact let
us assume for simplicity that $-g_{u}m_{d}A_{1/2}^{\Phi}(\tau_{u})\tan\beta=g_{d}m_{u}A_{1/2}^{\Phi}(\tau_{d})$ for all vector-like quarks and
leptons (with $u\to \nu$ and $d\to l$). In such a case the vector-like fermion contribution to the gluon fusion amplitude production for the 125 GeV
Higgs vanishes in the alignment limit, according to eq.~\eqref{Agg} for $\Phi=h$. On the other hand, the vector-like fermion
contribution to the $h\to\gamma\gamma$ amplitude vanishes if:
\begin{equation}
 N_c^u Q_{u}^2 - N_c^d Q_{d}^2 + Q_{\nu}^2 - Q_{l}^2=0 \,.
\end{equation}
Interestingly, the above condition is fulfilled if vector-like fermions have the same pattern of charges as the SM fermions.

In our numerical examples we fix $-g_{u}\tan\beta=g_d=1$ for all vector-like fermions. There are two important consequences of using this relation.
First: the 125 GeV Higgs production is exactly the same as in the SM. Second: for moderate and large $\tan\beta$ couplings of all Higgses to
up-type vector-like fermions are suppressed. We also take, for simplicity, all vector-like quarks masses and leptons equal to $m_{VLQ}$
and $m_{VLL}$, respectively. If, in addition, $m_{VLL}=m_{VLQ}$ the $h\to\gamma\gamma$ rate is also exactly the same as in the SM. However, even if
vector-like quarks are not degenerate with vector-like leptons the $h\to\gamma\gamma$ rate is still in good agreement with the LHC Higgs data
\cite{Higgsdata} if $\tan\beta$ is not small. This follows from the fact that  $h$ couplings to up-type (down-type)
vector-like fermions are suppressed by $g_{u}$ ($\cos\beta$) and only one family of vector-like fermions is introduced to explain the 750 GeV excess.
Notice also that the condition $-g_{u}\tan\beta=g_d$ implies that for moderate and large $\tan\beta$ only down-type
vector-like fermions give non-negligible contribution to the gluon fusion and $\gamma\gamma$ amplitudes for $H$ and $A$. 

\begin{figure}[t]
\center
\includegraphics[width=0.51\textwidth]{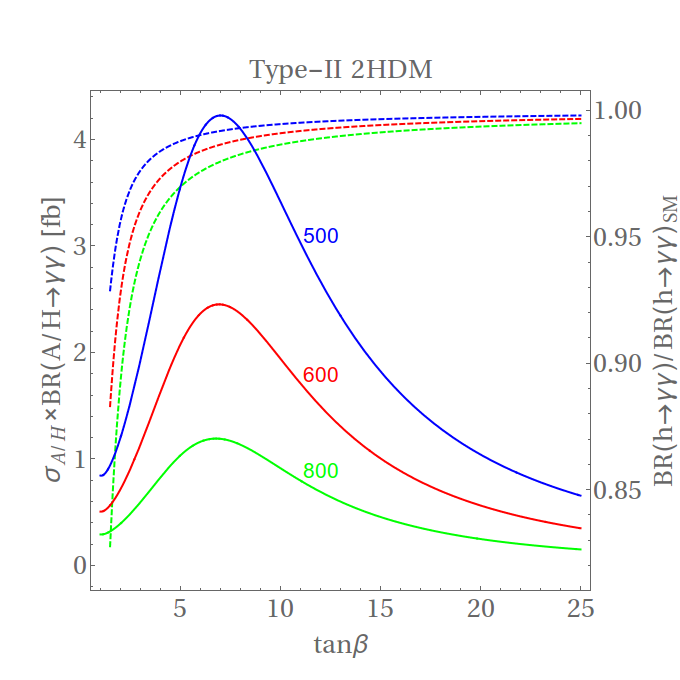}
\includegraphics[width=0.45\textwidth]{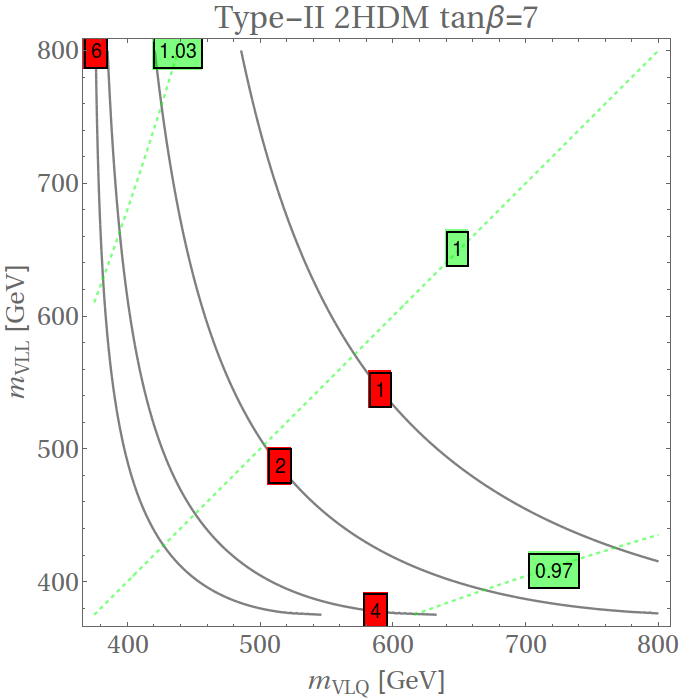}
\caption{$\sigma_{H/A}\times {\rm BR}(H/A\to\gamma\gamma)$ (solid lines) and ${\rm BR}(h\to\gamma\gamma)$ normalized to SM (dashed lines) in
Type-II 2HDM with one family of vector-like fermions \eqref{vectorlikeset} with
$-g_{u}\tan\beta=g_d=1$. In the left panel, dependence on $\tan\beta$ is shown for $m_{VLL}$=400 GeV and blue, red, green lines (from top to
bottom)
correspond to $m_{VLQ}=$ 500, 600, 800 GeV, respectively. In the right panel, $\tan\beta=7$ while $m_{VLL}$ and $m_{VLQ}$ are varied. The numbers on
solid (dashed) contours in the red (green) square boxes correspond to $\sigma_{H/A}\times {\rm BR}(H/A\to\gamma\gamma)$ in fb (${\rm
BR}(h\to\gamma\gamma)$ normalized to SM). 
}
\label{fig:x750_II}
\end{figure}

In the left panel of Fig.~\ref{fig:x750_II} we present dependence of the sum of diphoton signal rates from $H$ and $A$ decays, $\sigma_{H/A}\times
{\rm BR}(H/A\to\gamma\gamma)$, on $\tan\beta$ for $m_{VLL}=400$ GeV and several values of $m_{VLQ}$. We assume that $H$ and $A$ are degenerate  with
mass of 750 GeV. Note that due to particular values of form-factors the diphoton signal from $A$ decays is larger by a factor of five or more than
that from $H$ decays.  It can be seen that the 750 GeV diphoton signal is much larger for $\tan\beta$ around 7 than for small $\tan\beta$ and can be
of order $\mathcal{O}(1)$ fb for $m_{VLQ}=800$ GeV. In order to get 4 fb one needs $m_{VLQ}\sim 500$ GeV. The latter values may be in tension with
the LHC constraints for vector-like quarks \cite{LHC_vectorlike}, which are, however, model dependent and it is beyond the  scope of the present paper
to investigate them in detail. Note, also, that $m_{VLQ}$ can be larger for larger values of $g_d$. Notice also that despite the fact that
vector-like quarks are not degenerate with vector-like leptons the diphoton signal of the 125 GeV Higgs is very close to the SM prediction for
moderate and large $\tan\beta$, as explained before. In the right panel of Fig.~\ref{fig:x750_II} we
present the 750 GeV diphoton signal in the plane $m_{VLQ},m_{VLL}$ for optimal value $\tan\beta=7$. It can be seen, in particular, that lowering
$m_{VLL}$ to 375 GeV, which is a minimal value for which $H/A$ decays to vector-like fermions may not increase the total decay width, allow for
increase of $m_{VLQ}$ by about 150 GeV keeping the same cross-section and Yukawa couplings. Notice also that for this value of $\tan\beta$ deviations
from the SM prediction for the $h\to\gamma\gamma$ rate are at the level of few percent at most. Of course, in order to relax requirements on the
masses of
vector-like fermions and Yukawa couplings one can include additional copies of vector-like fermions \eqref{vectorlikeset} or to use bigger charges for
vector-like quarks and/or leptons. However, in the latter case one should keep in mind that the production and/or decays of the 125 GeV Higgs might be
affected.

\begin{figure}[t]
\center
\includegraphics[width=0.51\textwidth]{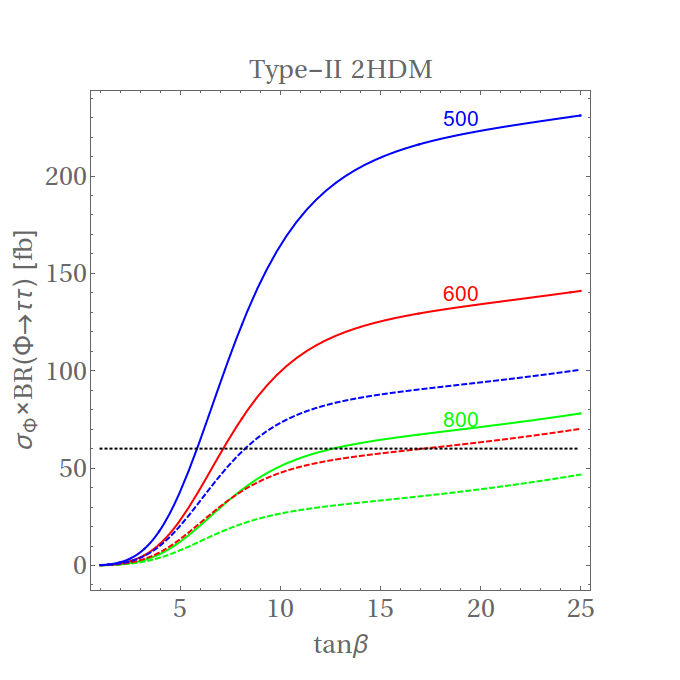}
\caption{$\sigma_{A}\times {\rm BR}(A\to\tau\tau)$ (solid lines) and $\sigma_{H}\times {\rm BR}(H\to\tau\tau)$ (dashed lines) in
Type-II 2HDM with one family of vector-like fermions \eqref{vectorlikeset} with
$-g_{u}\tan\beta=g_d=1$ as a function of $\tan\beta$. Blue, red, green
lines (from top to bottom)
correspond to $m_{VLQ}=$ 500, 600, 800 GeV, respectively. Horizontal
black dotted line corresponds to the experimental upper bound from ATLAS \cite{LHC_Atautau}.
}
\label{fig:xtautau}
\end{figure}

Let us also discuss constraints on this scenario from direct searches for heavy Higgs bosons in the $\tau\tau$ final state performed at the LHC. An
upper limit for the production cross-section times $\tau\tau$ branching fraction of a 750 GeV scalar boson at 13 TeV is about 60 fb
\cite{LHC_Atautau}.\footnote{Even though $b\bar{b}$ branching fraction of $H/A$ is larger than the $\tau\tau$ one, the $b\bar{b}$ channel is
experimentally much more challenging so constraints from $\tau\tau$ channel are stronger.} In Fig.~\ref{fig:xtautau} we present dependence of the 
$\tau\tau$ signal rates from $H$ and $A$ decays on $\tan\beta$ for several values of $m_{VLQ}$. It can be seen that $\tan\beta$ is constrained from
above by the $\tau\tau$ search. The constraint on $\tan\beta$ is stronger for lighter vector-like quarks because this makes the gluon fusion
production cross-section of heavy Higgses larger. Nevertheless, even for $m_{VLQ}=500$ GeV values of $\tan\beta\lesssim 5$, which correspond to
the diphoton signal of up to 4 fb (cf. Fig.~\ref{fig:x750_II}), are allowed by the current data. The tension between the diphoton signal and the
constraints  from the $\tau\tau$ search can be relaxed by reducing $H/A$ production cross-section while increasing branching fraction to diphoton
which can be realized, for example, by taking the heavy Higgs couplings to vector-like leptons larger than those to vector-like quarks. In any case
the interesting part of parameter space will be probed in near future by searches in the $\tau\tau$ channel.

\begin{figure}[t]
\center
\includegraphics[width=0.48\textwidth]{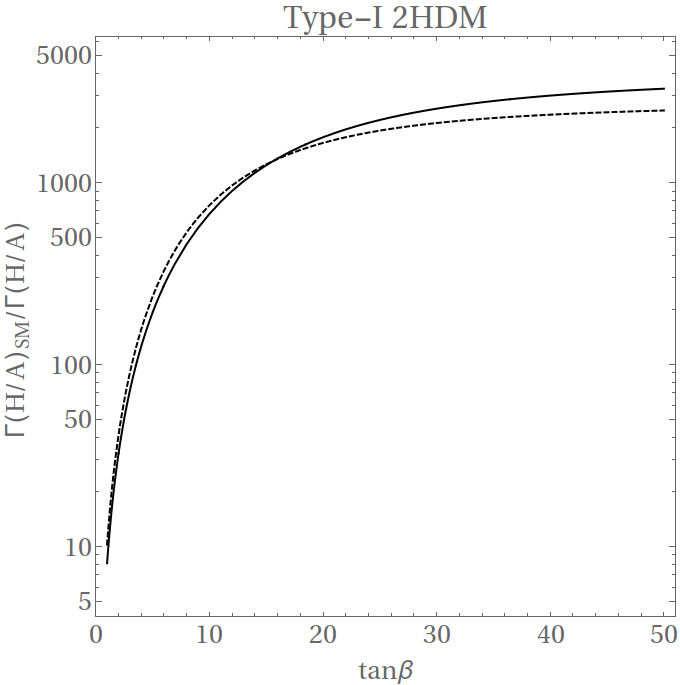}
\includegraphics[width=0.48\textwidth]{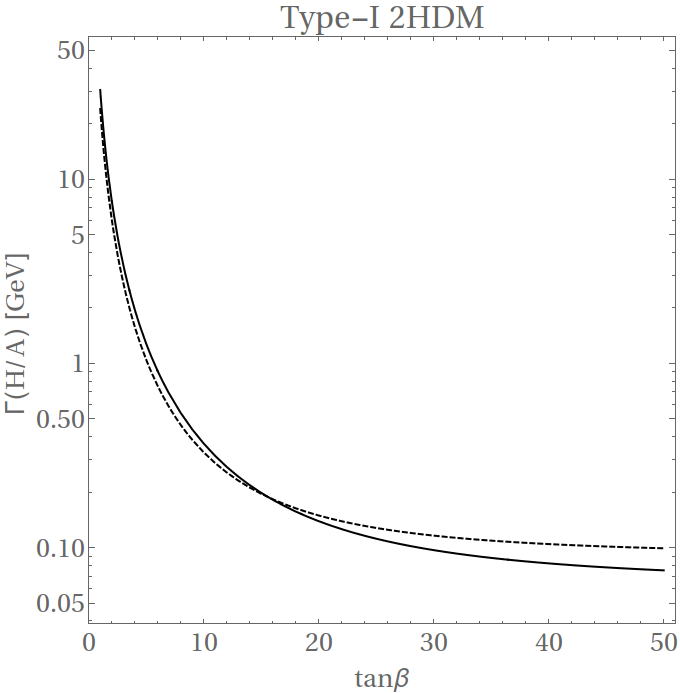}
\caption{The same as in Fig.~\ref{fig:decaywidth_II} but for Type-I 2HDM. $\Gamma(H/A\to gg)$ is fixed to the SM value. Difference between the
CP-even and CP-odd Higgses comes from a different phase space suppression in $H/A\to t\bar{t}$
(significant for smaller $\tan\beta$) and different form factors in the  $\Gamma(H/A\to gg)$ amplitude (important for large $\tan\beta$). 
}
\label{fig:decaywidth_I}
\end{figure} 

\subsection{Type-I 2HDM}

\begin{figure}[t]
\center
\includegraphics[width=0.51\textwidth]{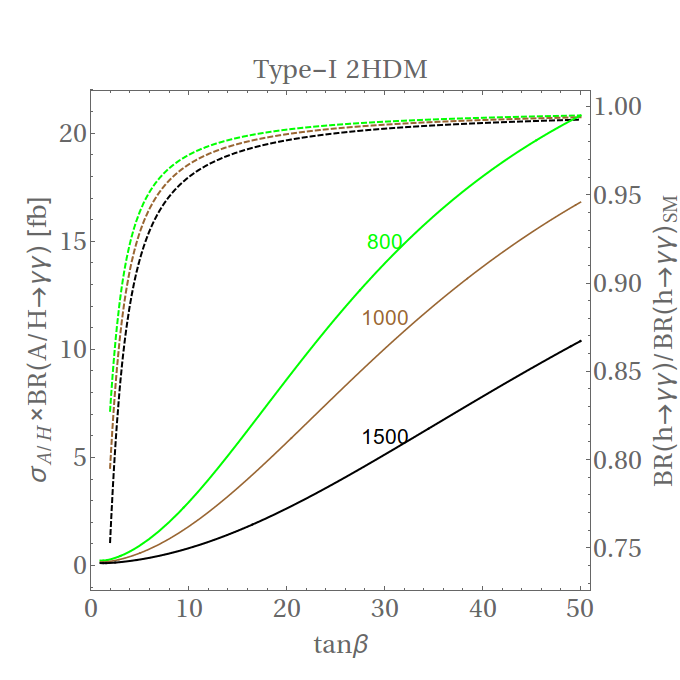}
\includegraphics[width=0.45\textwidth]{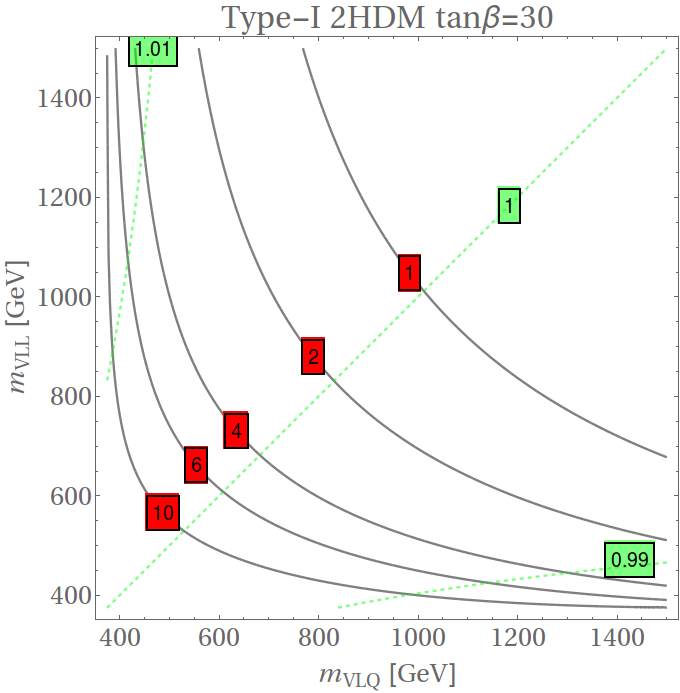}
\caption{The same as in Fig.~\ref{fig:x750_II} but in Type-I 2HDM. In the left panel, $m_{VLL}$=400 GeV and
 green, brown, black lines (from top to bottom)
correspond to $m_{VLQ}=$ 800, 1000, 1500 GeV, respectively. In the right panel, $\tan\beta=30$ is fixed.
}
\label{fig:x750_I}
\end{figure}

Let us now consider Type-I 2HDM in which the $H/A$ couplings to bottom quarks are scaled by $1/\tan\beta$, similarly as the corresponding couplings
to top quarks. In consequence, the total decay width of $H/A$ does not have a minimum as a function of $\tan\beta$, as can be seen from
 Figure~\ref{fig:decaywidth_I}. For very large values of $\tan\beta$ the total decay width of $H/A$ tends to $\Gamma(h\to gg)$. For
the SM 750 GeV Higgs $\Gamma(H\to gg)\approx0.06$ GeV corresponding to BR$(H\to gg)\approx2.5\times10^{-4}$ which means that for strongly suppressed
top quark Yukawa coupling the total decay width can be suppressed by a factor of 4000. Due to larger form factor for $A$ for strongly suppressed top
quark
Yukawa coupling the total decay width of $A$ is suppressed by about 2700. Suppressed top quark Yukawa
coupling leads to even stronger suppression of $\Gamma(h\to gg)$. However, in order to have large enough $H/A$ production cross-section to explain the
750 GeV excess new coloured particle must enhance the $H/A$ effective coupling to gluons to a similar level as the top quark loop does in the SM. In
the narrow width approximation, that we use throughout this paper and is fully justified, $\sigma(gg\to H/A) \sim\Gamma(H/A\to gg)$ so one should not
expect $\Gamma(H/A\to gg)$ to be smaller than $\mathcal{O}(0.01)$ GeV. Assuming the SM value for $\Gamma(H/A\to gg)$, the total decay width vary most
rapidly up to $\tan\beta\approx20$ for which $\Gamma(H/A\to t\bar{t})\approx\Gamma(H/A\to gg)$.  

In order to demonstrate consequences for the 750 GeV diphoton signal we choose the same model for vector-like fermions as for the Type-II 2HDM. The
results are shown in Figure~\ref{fig:x750_I}. From the left panel it can be seen that the 750 GeV diphoton rate increases indefinitely with
$\tan\beta$. Moreover, the diphoton signal can have correct magnitude to fit the 750 GeV excess, without invoking very large Yukawa couplings or
small masses for vector-like quarks. For example in the case of $\tan\beta=30$, presented in the right panel of Figure~\ref{fig:x750_I} with the same
assumptions about Yukawa couplings as in the Type-II 2HDM examples, masses of vector-like quarks can be above 1 TeV even if the vector-like lepton
masses are far away from the kinematic threshold and $H/A\to\gamma\gamma$ decays are not enhanced by a  large value of the form factor. Moreover,
the $h\to\gamma\gamma$ rate is within one percent from the SM prediction.

\section{Conclusions}
\label{sec:concl}

We have investigated a possibility that a tentative 750 GeV diphoton excess reported by ATLAS and CMS is the first signal of heavier Higgs bosons in
2HDMs. While it is not possible to fit this excess in a pure 2HDM, it is possible to do it when new particles are coupled to the Higgs sector. For
$\tan\beta\sim1$, even in the alignment limit, large multiplicity of new states with exotic electromagnetic charges are preferred to fit the excess.
Apart from aesthetic arguments, larger multiplicities of states are more likely to affect the production and decays of the 125 GeV, that are subject
to strong LHC constraints, thus complicating model building. In order to avoid large multiplicity of new particles, small total decay width is
preferred. In the context of 2HDM, the total decay width is suppressed for $\tan\beta$ significantly above one, due to suppression of the top Yukawa
coupling. In the Type-II 2HDM, the biggest suppression of the total decay width, as compared to the SM,  is about 250 which is obtained for
$\tan\beta$ around 7. In the Type-I 2HDM, the total decay width decreases monotonically with $\tan\beta$, approaching for very large $\tan\beta$ the
decay width into gluons which is typically few times $10^{-4}$ smaller than the total decay width of the SM 750 GeV Higgs. 

Due to large suppression of the total decay width it is possible to fit the 750 GeV excess with a small number of new particles. However, in contrast
to small $\tan\beta$ case, at least one of these particles must carry colour charge, otherwise gluon fusion cross-section would be strongly suppressed
due to smallness of the top Yukawa coupling. As a proof of concept, we have shown that adding to 2HDMs one family of vector-like quarks and leptons
with the corresponding SM fermion charges  is enough to fit the 750 GeV excess. Moreover, for such choice of vector-like fermions charges their
total contribution to the 125 GeV Higgs signal rates can vanish.  While in the Type-II model new fermions must have relatively large
Yukawa couplings and masses close to the experimental bounds, in the Type-I model parameters are not strongly constrained provided that $\tan\beta$
is large enough. We should emphasize that the 750 GeV excess is expected to be fitted also in many other extensions of 2HDMs without introducing large
multiplicities of new states. 

In the regions of $\tan\beta$ considered in this paper the total decay width of $H/A$ is around or below 1 GeV. The ATLAS 13 TeV data shows some
preference for much larger width of about 30 GeV. Even though CMS and 8 TeV data do not support this interpretation it is worth pointing out that
in the presented scenario single wide resonance preferred by the ATLAS 13 TeV data can be mimicked by $H$ and $A$ with masses that differ by few
tens of GeV. In such a case $H$ and $A$ contribute to different bins in the ATLAS analysis improving the
fit to the ATLAS 13 TeV data. Nevertheless, if the diphoton signal is real future LHC data will discriminate this hypothesis against
single wide resonance.

If the future LHC data confirm that the 750 GeV diphoton excess is due to a new resonance one of the next steps will be to measure its CP properties.
In 2HDM the diphoton signal from CP-odd Higgs decays is stronger than from the CP-even one. Nevertheless, CP-even state can by its own explain the
excess, which is especially simple in extensions of the Type-I 2HDM.

\section*{Acknowledgments}
This work has been partially supported by National Science Centre under research grants
DEC-2012/05/B/ST2/02597 and DEC-2014/15/B/ST2/02157. MB acknowledges support from the Polish 
Ministry of Science and Higher Education (decision no.\ 1266/MOB/IV/2015/0).
MB would like to thank Carlos Wagner for useful discussions.


\end{document}